\documentclass[english,twocolumn,floatfix,prl,aps,showpacs,superscriptaddress,bibliography]{revtex4-2}
\usepackage[english]{babel}                                                           
\usepackage[utf8]{inputenc}                                                            
\usepackage[T1]{fontenc}
\usepackage{epsfig, mathtools}
\usepackage{amsmath,dsfont}
\usepackage{hyphenat}
\usepackage{hyperref}
\usepackage{placeins}

\def\be{\begin{equation}}
\def\ee{\end{equation}}
\def\bea{\begin{eqnarray}}
\def\eea{\end{eqnarray}}
\def\ba{\begin{array}}
\def\ea{\end{array}}

\begin{document}

\title{Thermo-electric response in two-dimensional Dirac systems:\\ the role of particle-hole pairs}

\author{Kitinan Pongsangangan}
\author{Simonas Grubinskas}
\author{Lars Fritz}
\affiliation{Institute for Theoretical Physics, Utrecht University, 3584 CE Utrecht, The Netherlands}


\begin{abstract}
\noindent Clean two-dimensional Dirac systems have received a lot of attention for being a prime candidate to observe hydrodynamical transport behavior in interacting electronic systems. This is mostly due to recent advances in the preparation of ultrapure samples with sufficiently strong interactions. In this paper, we investigate the role of collective modes in the thermo-electric transport properties of those systems. We find that dynamical particle-hole pairs, plasmons, make a sizeable contribution to the thermal conductivity. While the increase at the Dirac point is moderate, it becomes large towards larger doping. We suspect, that this is a generic feature of ultraclean two-dimensional electronic systems, also applicable to degenerate systems. 
\end{abstract}
\maketitle


The study of transport in metallic systems has a long and successful history, going back all the way to Drude transport theory \cite{Drude1900}. Remarkably, a picture of non-interacting electrons scattering from disorder provides a reasonably description of transport properties of metals or more specifically Fermi liquids. In general, however, electrons are interacting and a question that has been discussed for 70 years is why the picture of independent electrons diffusing in a disordered background is so successful \cite{Ashcroft&Mermin1976}.  To rephrase the question:  Why and how do the long-range correlations between electrons due to Coulomb interaction become ineffective? Bohm and Pines argued that there are two components associated with the Coulomb interaction, short- and long-range, and they play a very different role. The short-range part leads to a quasiparticle renormalization in the spirit of Landau's Fermi liquid theory, leading to 'new' almost free electrons. The primary manifestation of the long-range part are the plasma oscillations or plasmons \cite{Pines&Bohm1952,Bohm&Pines1953}. 
In the standard theory of metallic transport, interactions consequently play a minor role in the low temperature limit. There are two ways in which they enter: (i) as a source of inelastic scattering for the electronic quasiparticles \cite{Landau1965} and (ii) as dynamical collective degrees of freedom like plasmons that make a direct contribution to transport properties. In conventional three dimensional Fermi liquids, neither of the two happens. (i) Inelastic scattering is subdominant compared to elastic impurity scattering. It is parametrically small in $(T/T_F)^2$ where $T_F \approx 10^3-10^5$ K is the Fermi temperature of the metallic system. (ii) In a three dimensional metal, stable plasmons are gapped, showing a gap that is even larger than the Fermi energy of the electronic system. This implies that they cannot be excited at energy scales relevant for transport~\cite{Pines1953}.  As a consequence, only electrons are relevant in the low-energy limit, and they interact with each other through the residual short-range component of the interaction. The primary source of scattering is given by disorder (note that we do not consider the role of phonons throughout this work \cite{Ziman}). 
One of the consequences of this is the famous Wiedemann-Franz law which goes back to 1853 \cite{WiedemannFranz}. It states that at lowest temperatures in metals the ratio
\begin{eqnarray}\label{eq:WF}
\lim_{T\to 0} \frac{\kappa}{T\sigma}=L_0
\end{eqnarray}
is constant and independent of details of the system. 
In Eq.~\eqref{eq:WF} $T$ is the temperature, $\sigma$ is the electrical conductivity, $\kappa$ is the heat conductivity, and $L_0=(\pi k_B)^2/(3e^2)$ is the Lorenz number ($k_B$ is the Boltzmann constant and $e$ the electron charge).
One way to rationalize this finding is that at lowest energies only electrons carry charge and heat, and both transport channels undergo the same scattering mechanisms from disorder. This leads to the same scattering time for both charge and heat transport. 
The question whether inelastic scattering can be the dominant scattering mechanis in degenerate Fermi systems has been discussed in the 1960s~\cite{Gurzhi,Abrikosov,PinesNozieres} but recently gained more momentum~\cite{Molenkamp,LucasSarma,Vignale,Benia,LevchenkoSchmalian,phononLucas}. The general expectation is that one should then observe hydrodynamic transport phenomena.

Another variant of conducting and interacting electronic systems are Dirac- and/or Weyl-metals~\cite{RevModPhys.90.015001}. Their defining feature is a linear band crossing in isolated points in the Brillouin zone which strongly suppresses the density of states. These systems are semimetals or non-degenerate. The most famous example is graphene which has been at the forefront of research for almost two decades~\cite{Novoselov2005,RevModPhys.81.109}. 
 
Close to its Dirac point, pristine graphene has properties that are distinct from normal Fermi liquids. One major difference is that at the Dirac point the system is scale-free, resembling a quantum critical system~\cite{Sheehy2007}. Consequently, temperature T is the only energy scale, contrary to a degenerate fermionic system which possesses the Fermi temperature $T_F$. While this modifies thermodynamic properties, it also has consequences on the interaction properties: inelastic interaction scattering  cannot be suppressed by the smallness of $T/T_F$ (also in the vicinity of the Dirac point $T/T_F$ remains large), it can even dominate elastic scattering from disorder. Therefore, in sufficiently clean samples, it is theoretically expected that one finds hydrodynamic transport behavior~\cite{Hartnoll2007,HartnollMueller2007,Kashuba2008,Fritz2008,MuellerFritz2008,Foster2008,Schuett2011,HydroNarozhny2015,Narozhny2017,Narozhny2019}. 
Secondly, it is known that decreasing dimensionality increases the effect of long-range interactions in electronic systems. That not only increases inelastic scattering, it also makes the effect of collective modes more prominent. It is known that plasmons in a one-dimensional conductor contribute significantly to thermal transport. This begs the question about two dimensions. In two dimensions, contrary to three dimensions, plasmons are massless [5, 6, 7] and follow a square root dispersion, {\it i.e.}, $\omega \propto \sqrt{q}$. Consequently, they are easily excited under non-equilibrium conditions and can therefore be relevant to the transport phenomena, especially thermal transport. It is important to note that the second point is not exclusive to Dirac systems but it is also true for two-dimensional degenerate systems.

In recent years, suspended samples or samples sandwiched in between boron-nitrid (BN) structures~\cite{Elias2011,Crossno1058,Lucas2016,Bandurin1055,Bandurin2018,Braem2018,Sulpizio2019,Berdyugin162,Ella2019,Gallagher158} allow to suppress disorder levels sufficiently to access the hydrodynamic regime. 
With the new ultrapure samples, it is thus possible to ask quantitative questions that could not be addressed before. In this paper, we reinvestigate transport theory in ultraclean Dirac systems. Our special focus is on the role of Coulomb interactions and their unscreened long-range nature in thermo-electric transport.

As explained above, we expect Coulomb interaction to be responsible for mainly two effects in regards of transport phenomena: (i) charge carriers scatter from each other leading to an effective inelastic transport time or mean free path; (ii) collective excitations, such as plasmons, that possess their own dynamics. Consequently, they make a direct contribution to the heat current. 

\begin{figure}[h]
\includegraphics[width=0.45\textwidth]{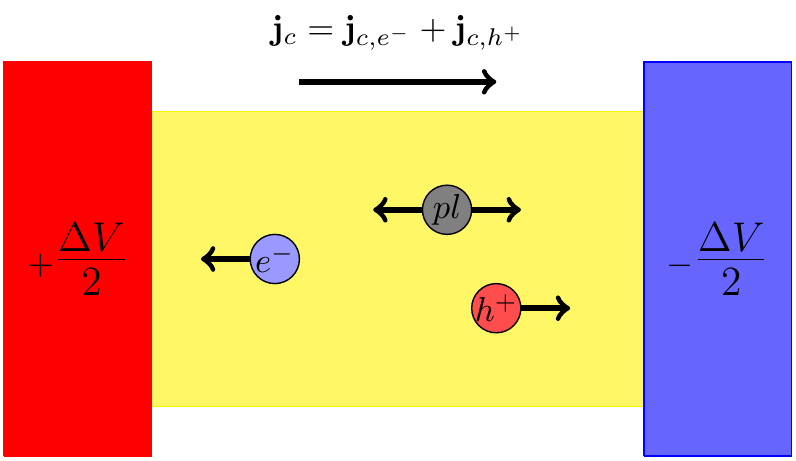}
\caption{Electrons and holes react oppositely to an applied voltage drop. Plasmons do not couple directly to the voltage difference, however, they experience drag and serve as a source of inelastic scattering.}\label{fig:electriccurrent}
\end{figure}
The thermo-electric response involves two types of currents: the electrical current $J^e$ and the heat current $Q=J^E-\mu/eJ^e$, where $J^E$ is the energy current. The Onsager relation states~\cite{Mahan} 
\begin{eqnarray}
\left(\begin{array}{c} \vec{J}^e \\ \vec{Q} \end{array} \right)	=\left(\begin{array}{cc} \hat{\sigma} & \hat{\alpha} \\ T \hat{\alpha} & \hat{\overline{\kappa}} \end{array} \right) \left(\begin{array} {c} \vec{E}\\ -\vec{\nabla}T \end{array} \right)\;.
\end{eqnarray}
The thermal conductivity, $\hat{\kappa}$, is defined as the heat current response to a thermal gradient $-\vec{\nabla}T$ in the absence of an electrical current (electrically isolated boundaries), given by $\hat{\kappa}=\hat{\overline{\kappa}}-T\hat{\alpha}\hat{\sigma}^{-1}\hat{\alpha}$. In the following we drop 'hats' and only explicitly discuss the diagonal response $\sigma$ and $\kappa$. The Wiedemann-Franz ratio $\kappa/(T \sigma)$ assumes the value $L_0=\pi^2/3 \times(k_B/e)^2$ ($L_0$ is the Lorenz number) in a Fermi liquid~\cite{WiedemannFranz}, see Eq.~\eqref{eq:WF}. This is sometimes considered the hallmark of a Fermi liquid and it was argued before that it breaks down in the vicinity of the Dirac point~\cite{muellerhydro2018,MuellerFritz2008,Lucas2016}. The main reason for this breakdown is that there are two completely independent hydrodynamic modes that are subject to different scattering mechanisms.

\begin{figure}
\includegraphics[width=0.45\textwidth]{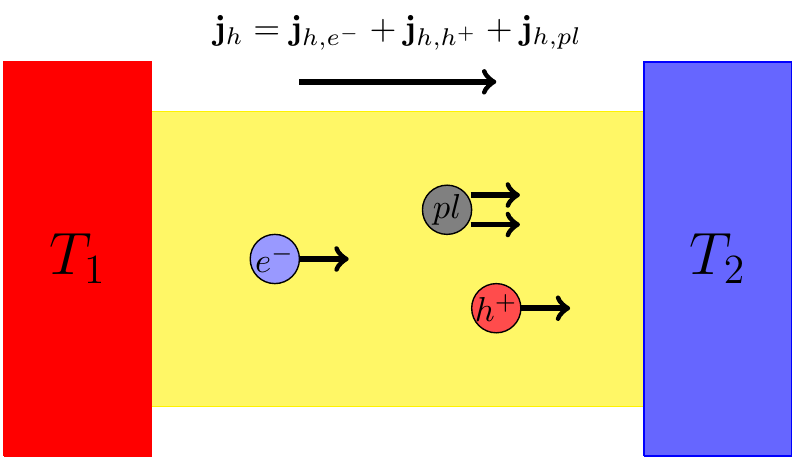}
\caption{Electrons, holes, and plasmons react to a temperature gradient in the same manner. Plasmons make a direct contribution to the heat current.}\label{fig:heatcurrent}
\end{figure}
Instead of concentrating on two types of degrees of freedom, electrons and holes, we additionally consider plasmons. 
The situations are sketched in Fig.~\ref{fig:electriccurrent} for electrical transport and in Fig.~\ref{fig:heatcurrent} for heat transport. The potential drop in Fig.~\ref{fig:electriccurrent} acts on electrons and holes oppositely, but not on the plasmons: the plasmons  experience no direct force, but they are subject to drag effects. Being neutral quasiparticles, they do not contribute to the charge current. In the case of a temperature gradient, see Fig.~\ref{fig:heatcurrent}, all particles experience a force in the same direction, and there is an additional direct contribution to the current through the plasmons. On a technical level, we derive and solve three coupled Boltzmann equations for electrons, holes, and plasmons, which includes relaxational processes and the respective streaming terms. 

\noindent{\it{Main result}:} We find that the plasmon contribution to the heat conductivity is seizable and cannot be neglected, neither at the Dirac point, nor in the degenerate limit.

\noindent {\it{The model:}} We study a model of Dirac fermions coupled through Coulomb interaction and subject to potential disorder:
\begin{eqnarray}\label{eq:model}
H&=& \int d^2 \vec{r} \;  \Psi^\dagger_{i} (\vec{r}) \left(-i v_F \vec{\partial} \cdot \vec{\sigma} +V_{\rm{dis}}(\vec{r}) \right) \Psi^{\phantom{\dagger}}_{i} (\vec{r})  \nonumber \\ &+&\frac{1}{2}  \int d^2\vec{r} d^2 \vec{r}' \Psi^\dagger_i (\vec{r}) \Psi^{\phantom{\dagger}}_i (\vec{r}) V(\vec{r}-\vec{r}') \Psi^\dagger_j (\vec{r}') \Psi^{\phantom{\dagger}}_j (\vec{r}')\;.\nonumber \\
\end{eqnarray}
$\Psi_i(\vec{r})$ is the two component wave function, $i$ is the flavor index ranging from $i=1, ..., N$ (for graphene $N=4$ counting spin and valley), $v_F$ the Fermi velocity, and $V(\vec{r}-\vec{r}')=\alpha\frac{v_F}{|\vec{r}-\vec{r}'|}$ the Coulomb interaction (note that double indices are summed over). The dimensionless fine structure constant sets the strength of interaction and is given by $\alpha=e^2/(4 \pi \epsilon_0 \epsilon_r v_F)$. The disorder potential $V_{\rm{dis}} (\vec{r})$ can be used to describe a variety of disorder types specified by the disorder correlation function. We only consider delta-correlated disorder, {\it i.e.},  $\langle V_{\rm{dis}} (\vec{r}) V_{\rm{dis}} (\vec{r}') \rangle=4\pi \gamma^2 \delta (\vec{r}-\vec{r}')$, but generalizations are straightforward. Consequently, the parameters of our theory are $\alpha$ and $\gamma$.
Eq.~\eqref{eq:model} does not provide a convenient starting point for the study of the thermal conductivity due to the non-local Coulomb interaction. 

A corresponding local field theory that easily lends itself to an interpretation in terms of plasmons can be derived using a Hubbard-Stratonovich transformation. It reads
 \begin{eqnarray}\label{eq:action}
\mathcal{L}=&-&\frac{\epsilon_0 \epsilon_r}{2}  \phi (\vec{r},z,t)\left(\vec{\partial}^2+\partial_z^2 \right)\phi (\vec{r},z,t)\nonumber \\ &-&   e \Psi^\dagger_\alpha  (\vec{r},t) \Psi^{\phantom{\dagger}}_\alpha (\vec{r},t) \phi(\vec{r},z,t)\delta(z)	\\ &+&   \Psi^\dagger_\alpha (\vec{r},t)\left(i\partial_t-i v_F \vec{\partial}\cdot \vec{\sigma}+V_{\rm{dis}}(\vec{r§}) \right)\delta(z)\Psi^{\phantom{\dagger}}_\alpha (\vec{r},t)\;,\nonumber 
\end{eqnarray}
where $\phi (\vec{r},z,t)$ is the real valued plasmon field. Importantly, the mapping between Eq.~\eqref{eq:model} and Eq.~\eqref{eq:action} is exact.

\noindent{\it Current operators:} The electrical charge current is only carried by electrons and holes and its density is given by 
\begin{eqnarray}
{\bf{j}}^e (\vec{r},t)=-e v_F \Psi^\dagger_\alpha \left(\vec{r},t\right)\vec{\sigma}\Psi^{\phantom{\dagger}}_\alpha \left(\vec{r},t\right)\;.
\end{eqnarray}
The heat current density is given by ${\bf{Q}}={\bf{j}}^E -\mu/e {\bf{j}}^e$ where the energy current density reads~\cite{Peskin}
\begin{eqnarray}\label{eq:heatcurrent}
{\bf{j}}^E (\vec{r},z.t)	&=& -i v_F \Psi^\dagger_\alpha (\vec{r},t)\vec{\sigma}\partial_t\Psi^{\phantom{\dagger}}_\alpha (\vec{r},t) \nonumber \\ &+& \epsilon_0 \epsilon_r \vec{\partial} \phi(\vec{r},z,t) \partial_t \phi(\vec{r},z,t)\;.
\end{eqnarray}
This expression explicitly includes the plasmon contribution which is the main new aspect of this work.

\noindent{\it Plasmon dynamics:} Integrating out the photon modes outside the graphene sheet leads to an effective two-dimensional theory, $\mathcal{S}_{\Phi}=\frac{1}{2}\int dt dt' d^2 \vec{r} d^2\vec{r}' \Phi (\vec{r},t)D_0^{-1}(\vec{r},\vec{r}',t,t')\Phi(\vec{r}',t')$ with $\Phi(\vec{r},t)=\phi(\vec{r},z=0,t)$ and $D_0^{-1}(\vec{k},\omega)=\alpha (2\pi v_F)/(e^2 k)$ with $k=|\vec{k}|$. 

The dynamics is generated from within the fermionic system and corresponds to particle-hole pairs. We use the standard random phase approximation (RPA), formally justified in the limit of a large number $N$ of flavors. The boson self energy is approximated through $e^2 \Pi (\vec{r},\vec{r}',t,t')$, where $\Pi (\vec{r},\vec{r}',t,t')$ is the polarization function. The polarization function of two-dimensional Dirac systems has a closed analytical form at zero temperature~\cite{Wunsch_2006}. The finite temperature properties at arbitrary chemical potential have been studied numerically in Ref.~\cite{Sarmaplasmons}. In the long wavelength limit, important for the plasmon dynamics, the retarded polarization function can be approximated as
\begin{eqnarray}\label{eq:polfunc}
\Pi^r (\vec{q},\omega,\mu,T) &\approx &   \frac{Nq^2 T}{4\pi \omega^2}\ln \left(2+2\cosh\left(\frac{\mu}{T}\right) \right)\nonumber \\ &-&i \frac{Nq^2}{32 \omega} f(\mu,\omega,T)\;,
\end{eqnarray}
with $f(\mu,\omega,T)=2+\tanh \left(\frac{\mu}{2T}-\frac{\omega}{4T} \right) -\tanh \left(\frac{\mu}{2T}+\frac{\omega}{4T} \right)$. We obtain the plasmon dispersion from the poles of the retarded plasmon propagator 
\begin{eqnarray}
D^r(\vec{q},\omega)=\frac{D^r_0(\vec{q},\omega)}{1-e^2 D^r_0(\vec{q},\omega)\Pi^r(\vec{q},\omega,\mu,T)}\;,
\end{eqnarray}
 where $D_0^r(\vec{q},\omega)=1/(2 \epsilon_0 \epsilon_r q)$, with $\epsilon_0$ being the vacuum permittivity, while $\epsilon_r$ is the relative permittivity. Using the approximate polarization function, Eq.~\eqref{eq:polfunc}, we can approximate the plasmon propagator as 
\begin{eqnarray}
	D^r(\vec{q},\omega)\approx \frac{1}{2 \epsilon_0 \epsilon_r}\frac{\omega^2}{q}	\frac{1}{(\omega+i0^+)^2-\left(\omega_p(\vec{q})+i\gamma_p(\vec{q})\right)^2} \;, \nonumber \\
\end{eqnarray}
with the plasmon dispersion $\omega_p(\vec{q})$ and damping $\gamma_p(\vec{q})$ given by 
\begin{eqnarray}\label{eq:plasmondispersion}
\omega_p(\vec{q})&= & \sqrt{\alpha \frac{N}{2}k_B T v_F q \ln \left(2+2\cosh\left(\frac{\mu}{k_B T}\right)\right) }\;, \nonumber \\ \gamma_p(\vec{q})&= &- \frac{\pi \omega_p(\vec{q})^2}{16T} \frac{f(\mu,\omega_p(\vec{q}),T)}{\ln \left(2+2\cosh\left(\frac{\mu}{T}\right) \right)} \;.
\end{eqnarray}
There are two possible momentum cutoffs for the plasmons of Eq.~\eqref{eq:plasmondispersion}. Either, when they cease to be well-defined quasiparticles, {\it i.e.}, $\omega_p(\vec{q}_c) \approx \gamma_p(\vec{q}_c)$, or, when the square-root dispersion breaks down, {\it i.e.}, $q_c=\alpha N T/(2v_F)\ln \left(2+2\cosh \left(\mu/T \right) \right)$. Numerically, we always choose the lower of the two. In practice, it turns out that it is always the latter. For all practical calculations in this paper, we assume the plasmons to be well-defined quasiparticles since their decay rate is parametrically small in $\alpha$. It turns out that below $q_c$ plasmons are very stable against a single particle-hole decay channel. 
Therefore, the leading relaxation mechanisms might be from either the plasmon decay into two electron-hole pairs \cite{Kivenson1969} or phonon-assisted Landau damping \cite{Glazman2004}. Both are neglected in this work for different reasons. The former channel is of higher order in perturbation theory, while the latter is forbidden in the electron hydrodynamic window.
It is important to note that there is also a linear plasmon beyond the cutoff scale which is subleading and consequently negligible in our analysis.

\noindent{\it{The Boltzmann equation:}} We leave a systematic derivation of the Boltzmann equation starting from the Schwinger-Keldysh formalism (see for instance Ref.~\cite{kamenev_2011}) for the supplemental material. The key steps of the derivation are: (i) A conserving approximation of the fermion and boson self-energies to the lowest non-trivial order in $\alpha$ and $\gamma$. (ii) A lowest non-trivial order gradient expansion starting from the Wigner transform. (iii) an integration over the fermion and boson spectral functions, equivalent to an on-shell quasiparticle approximation; (iv) a projection into the quasiparticle basis. In the last step we only consider the diagonal parts and neglect Berry phase (these terms are second order in the gradient expansion) and Zitterbewegung terms (see Ref.~\cite{Kashuba2008,Fritz2008}). Eventually, we find three coupled Boltzmann equations for electrons, holes, and plasmons, 
\begin{eqnarray}\label{eq:Boltzmann}
&&e\vec{E} \partial_{\vec{k}} f_{\lambda}(\vec{k})-\lambda v_F \hat{\vec{k}}\sigma_z \vec{\nabla} T \partial_T f_\lambda(\vec{k})=I^\lambda_{\rm{coll}}[f_\lambda,b]\;, \nonumber \\&& 2\frac{\vec{k}}{k^2}\vec{\nabla}T \omega_p(\vec{k})\partial_T b_{\omega_p(\vec{k})}(\vec{k})=\tilde{I}_{\rm{coll}}[f_\lambda,b]\;.
\end{eqnarray}
Here, $f_\lambda$ and $b$ are the distribution functions of the electrons and holes ($\lambda=\pm$), and the plasmons, respectively. 

\noindent{\it{Sources of current relaxation}:}
The collision integral for the Dirac fermions consists of two independent parts,
\begin{eqnarray}
I^\lambda_{\rm{coll}}&=&\int \frac{d^2q}{(2\pi)^2}\sum_{\lambda'=\pm}\mathcal{C}^{\rm{inel}}_{\lambda \lambda'}(\vec{k},\vec{q}) \left[f_\lambda(\vec{k})\left(1-f_{\lambda'}(\vec{k}+\vec{q})\right)\right. \nonumber \\  &-& \left. b_{\epsilon_\lambda(\vec{k})-\epsilon
_{\lambda'}(\vec{k}+\vec{q})}(\vec{q})\left(f_{\lambda'}(\vec{k}+\vec{q})-f_\lambda(\vec{k})\right) \right] \nonumber \\ &+& \int \frac{d^2q}{(2\pi)^2} \mathcal{C}^{\rm{el}}_\lambda(\vec{k},\vec{q}) \left(f_\lambda(\vec{k})-f_{\lambda}(\vec{k}+\vec{q})\right)\;.
\end{eqnarray}
The first term accounts for inelastic scattering of electrons from plasmons, denoted $\mathcal{C}^{\rm{inel}}_{\lambda \lambda'}$. In this process, both energy and momentum are transferred between the fermions and the plasmons. Additionally, there is elastic scattering from disorder, encoded in $\mathcal{C}^{\rm{el}}_\lambda$. This term breaks momentum conservation and is important to relax the heat current. The collision integral for the plasmons reads
\begin{eqnarray}
\tilde{I}_{\rm{coll}}&=&\int \frac{d^2q}{(2\pi)^2}\sum_{\lambda,\lambda'=\pm}\tilde{\mathcal{C}}^{\rm{inel}}_{\lambda \lambda'}	\left[ f_{\lambda'}(\vec{k}+\vec{q})\left(f_\lambda(\vec{q})-1\right) \right. \nonumber \\  &-
& \left. b_{\epsilon_{\lambda'}(\vec{k}+\vec{q})-\epsilon_\lambda(\vec{q})}(\vec{k})\left(f_{\lambda'}(\vec{k}+\vec{q})-f_\lambda(\vec{q})\right) \right]\;.
\end{eqnarray}
It contains an inelastic part describing scattering from fermions, $\tilde{\mathcal{C}}^{\rm{inel}}_{\lambda \lambda'}$ (we defer the role of inelastic scattering from disorder to follow-up work).  In the absence of disorder, the combined electron-plasmon system conserves momentum. The precise form of $\mathcal{C}^{\rm{inel}}_{\lambda \lambda'}$, $\mathcal{C}^{\rm{el}}_\lambda$, and $\tilde{\mathcal{C}}^{\rm{inel}}_{\lambda \lambda'}$ can be found in the supplemental materials. It is important to note, however, that momentum excited in the plasmon sector can be relaxed in the fermion sector from disorder.

\noindent{\it{Linearized Boltzmann equation:}} In equilibrium, the collision integrals are nullified by the thermal Fermi-Dirac and Bose-Einstein distributions, respectively,
$f^0_\lambda(\vec{k})=	(e^{(\epsilon_\lambda(\vec{k})-\mu)/T}+1)^{-1}$ and $b^0_{\omega} (\vec{k})=(e^{\omega_p(\vec{k})/T}-1)^{-1}$.
In the presence of driving terms due to a potential gradient, a thermal gradient, or both, the distribution functions deviate from their equilibrium form.
Importantly, even though an electric field does not couple directly to the plasmons, away from the Dirac point they are still driven out of equilibrium by a drag effect~\footnote{At the Dirac point the underlying particle-hole symmetry forbids drag~\cite{Fritz2011} }. Since we are interested in linear response transport properties, we linearize the Boltzmann equations in $\vec{E}$ and $\vec{\nabla}T$. This enforces the following parametrizations for the fermions and boson distribution functions: 
\begin{widetext}
\begin{eqnarray}
f_\lambda(\vec{k})&=&f^0_\lambda(\vec{k})+1/T^2f^0_\lambda(\vec{k})\left(1-f^0_\lambda (\vec{k}) \right)\lambda v_F \hat{\vec{k}}	\cdot   \left(e\vec{E}  \chi^E_\lambda (k)+ \vec{\nabla}T   \chi^T_\lambda (k) \right)\; \quad{\rm{and}} \nonumber \\ 
 b_{\omega_p(\vec{k})} (\vec{k})&=& b^0_{\omega_p(\vec{k})}(\vec{k})+ 1/T^2 b^0_{\omega_p(\vec{k})}(\vec{k})\left( 1+ b^0_{\omega_p(\vec{k})}(\vec{k})\right) v_F \hat{\vec{k}} \cdot  \left(e\vec{E}  \phi^E (k)+ \vec{\nabla}T   \phi^T (k) \right)\;.
 \end{eqnarray}
 \end{widetext}
 As mentioned before, an electric field applied to the fermions 'drags' the plasmons out of equilibrium which is why we have to introduce $\phi^E$. The functions $\chi_\lambda^{T/E}$ and $\phi^{T/E}$ have to be determined numerically and give access to the respective currents and related response functions.
In terms of their parametrizations, we find the following set of equations 
\begin{widetext}
\begin{eqnarray}\label{eq:coupledboltz}
 	-\lambda e \frac{v_F}{T} \hat{\vec{k}}\cdot\vec{E}  f^0_{\lambda}(\vec{k})\left(1-f^0_{\lambda}(\vec{k}) \right)-\lambda \frac{v_F}{T} \hat{\vec{k}} \vec{\nabla} T \frac{\epsilon_\lambda(\vec{k})-\mu}{T} f^0_{\lambda}(\vec{k})\left(1-f^0_{\lambda}(\vec{k}) \right) &=& I_{\rm{inel}}^{\rm{lin}}\left[ \chi_T,\chi_E,\phi_T,\phi_E  \right]+I_{\rm{el}}^{\rm{lin}}\left[\chi_E,\chi_T\right] \nonumber \\ \frac{\vec{k}}{k^2}\vec{\nabla}T \frac{\omega_p^2(\vec{k})}{T^2}b^0_{\omega_p(\vec{k})}(\vec{k})\left( 1+ b^0_{\omega_p(\vec{k})}(\vec{k})\right)&=& \tilde{I}_{\rm{inel}}^{\rm{lin}}\left[ \chi_T,\chi_E,\phi_T,\phi_E  \right]+\tilde{I}_{\rm{inel}}^{\rm{el}}\left[ \phi_T,\phi_E  \right] \;,\nonumber \\
\end{eqnarray}
	\end{widetext}
with details to be found in the supplemental material. In the absence of disorder, the combined system of fermions and plasmons possesses a zero mode associated with momentum conservation, meaning the combination ${C}_{\lambda \lambda'}^{\rm{inel}}$ and $\tilde{C}_{\lambda \lambda'}^{\rm{inel}}$ together with the appropriate mode cannot relax momentum (we explicitly checked this point numerically). The intuition behind this is that momentum can always be transferred between the fermion and plasmon sector without being dissipated. This implies that disorder scattering is vital as a source of momentum relaxation for the total system, and it plays an important role in the choice of modes, explained below (see Ref.~\cite{Ziman} for a related discussion in the electron-phonon problem without Umklapp scattering).

\noindent{\it{Choice of modes:}} The above parametrization allows for a very transparent identification of the slow hydrodynamic modes of the problem. In the solution of the fermion only problem, Eq.~\eqref{eq:model}, it was pointed out that in order to study thermoelectric transport in the vicinity of the Dirac point all the way to the Fermi liquid regime it suffices to study two types of modes for electrons and holes, respectively, $\chi^{T/E}_\lambda (k)=a^{T/E}_{0,\lambda}+\lambda a^{T/E}_{1,\lambda} k$~\cite{MuellerFritz2008}. The mode associated with $a^{T/E}_{0,\lambda}$ is called chiral mode, whereas the one associated with $a^{T/E}_{1,\lambda}$ corresponds to the momentum mode.  For the plasmons, the equivalent ansatz reads $\phi^{T/E}(k)=b^{E/T}_0+b^{T/E}_1 k$. We can convert the problem of solving the Boltzmann equation into a linear algebra problem by projecting the scattering integral onto the respective modes, see the discussion in the supplemental material.  
\begin{center}
	\begin{figure}[h]
		\includegraphics[width=0.49\textwidth]{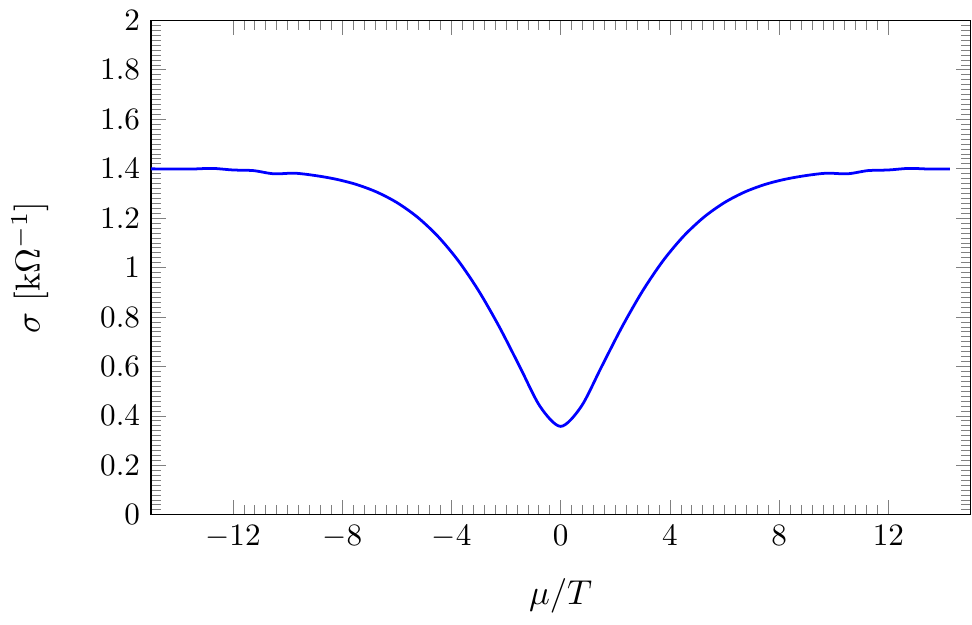}\caption{Electrical conductivity as a function of chemical potential. The curve was obtained as solution of the Boltzmann equation with $\alpha = 0.36$ and $4\pi \gamma^2 = 0.5$ at $T=75 K$.} \label{fig:electricalconductivitymu}
	\end{figure}
\end{center}
\begin{center}
	\begin{figure}[h]
		\includegraphics[width=0.5\textwidth]{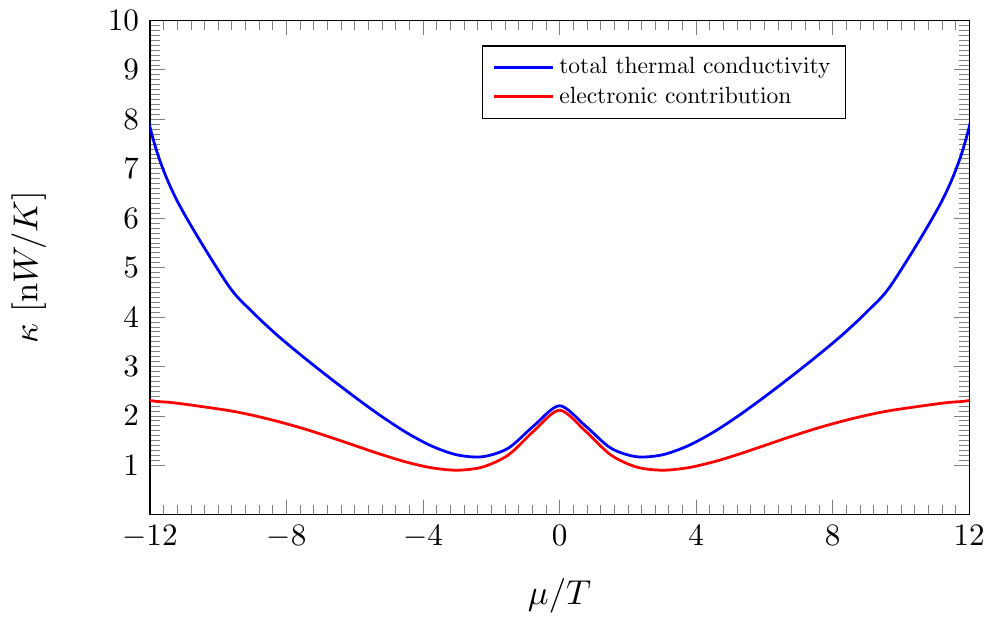}	
		\caption{Thermal conductivity of graphene as a function of the chemical potential. The curves are calculated using the same set of parameters used in Fig.~\ref{fig:electricalconductivitymu}. The blue curve is the full response including the plasmons while the red curve only shows the electronic contribution. There is a slight plasmon enhancement  close to the Dirac point and a massive one with increasing chemical potential.}  \label{fig:heatconductivitymu}
	\end{figure}
\end{center}

\begin{center}
\begin{figure}[h]
\includegraphics[width=0.48\textwidth]{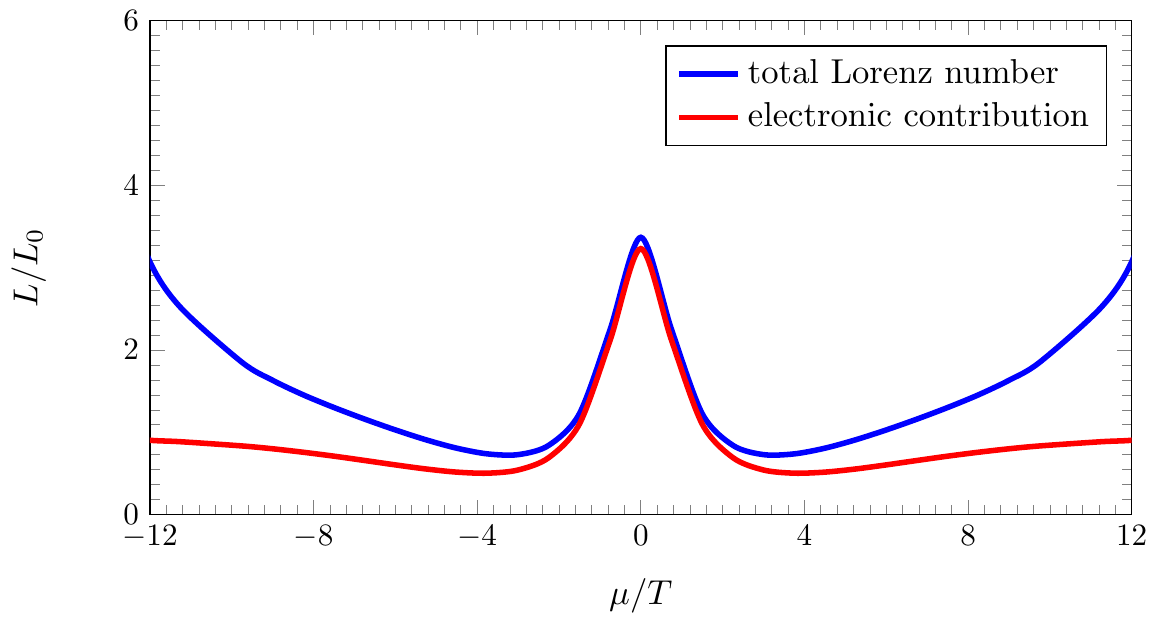}
\caption{Wiedemann-Franz ratio as a function of the chemical potential. We observe two regions with enhancement: the well-known hydrodynamic regime in the vicinity of the Dirac point and at higher doping. }\label{fig:WFmu}
\end{figure}
\end{center}

\begin{center}
	\begin{figure}[h]
		\includegraphics[width=0.49\textwidth]{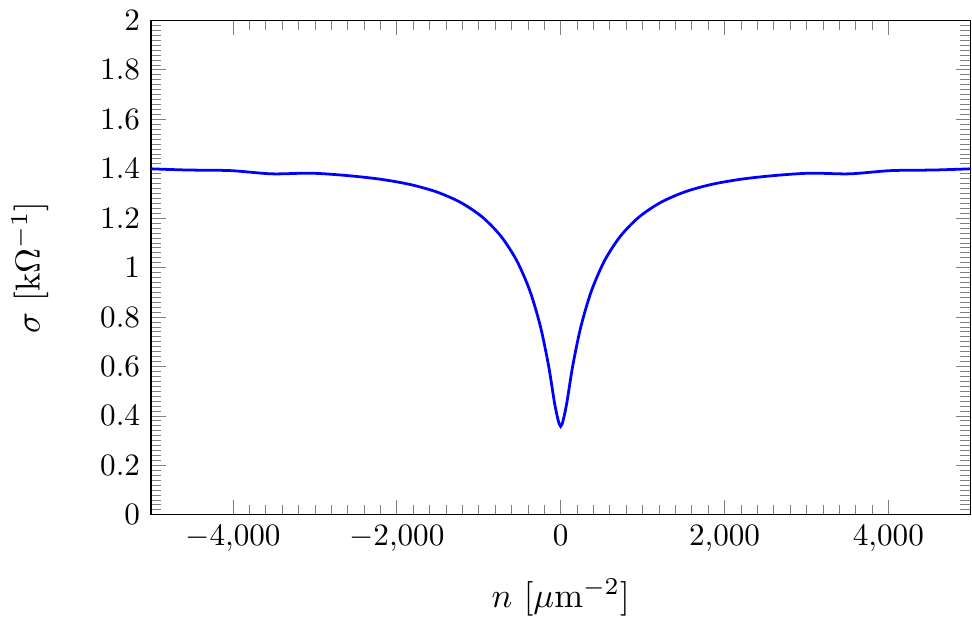}\caption{Electrical conductivity as a function of carrier density. The curve was obtained as solution of the Boltzmann equation with $\alpha = 0.36$ and $4\pi \gamma^2 = 0.5$ at $T=75 K$.} \label{fig:electricalconductivity}
	\end{figure}
\end{center}

\begin{center}
	\begin{figure}[h]
		\includegraphics[width=0.5\textwidth]{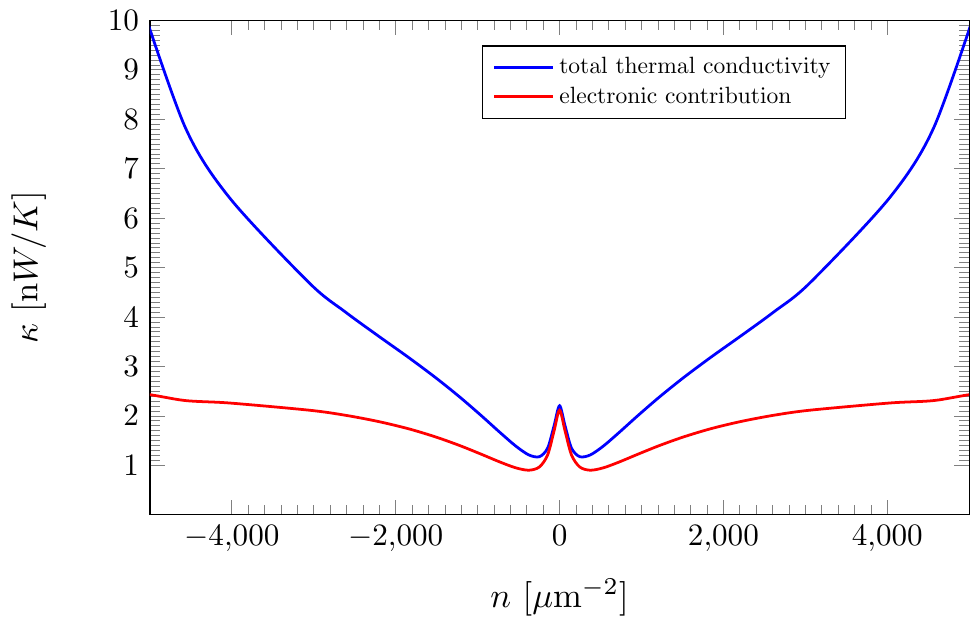}	
		\caption{Thermal conductivity of graphene as a function of the charge carrier density. The curves are calculated using the same set of parameters used in Fig.~\ref{fig:electricalconductivity}. The blue curve is the full response including the plasmons while the red curve only shows the electronic contribution. There is a slight plasmon enhancement  close to the Dirac point and a massive one with increasing chemical potential.}  \label{fig:heatconductivity}
	\end{figure}
\end{center}

\begin{center}
\begin{figure}[h]
\includegraphics[width=0.48\textwidth]{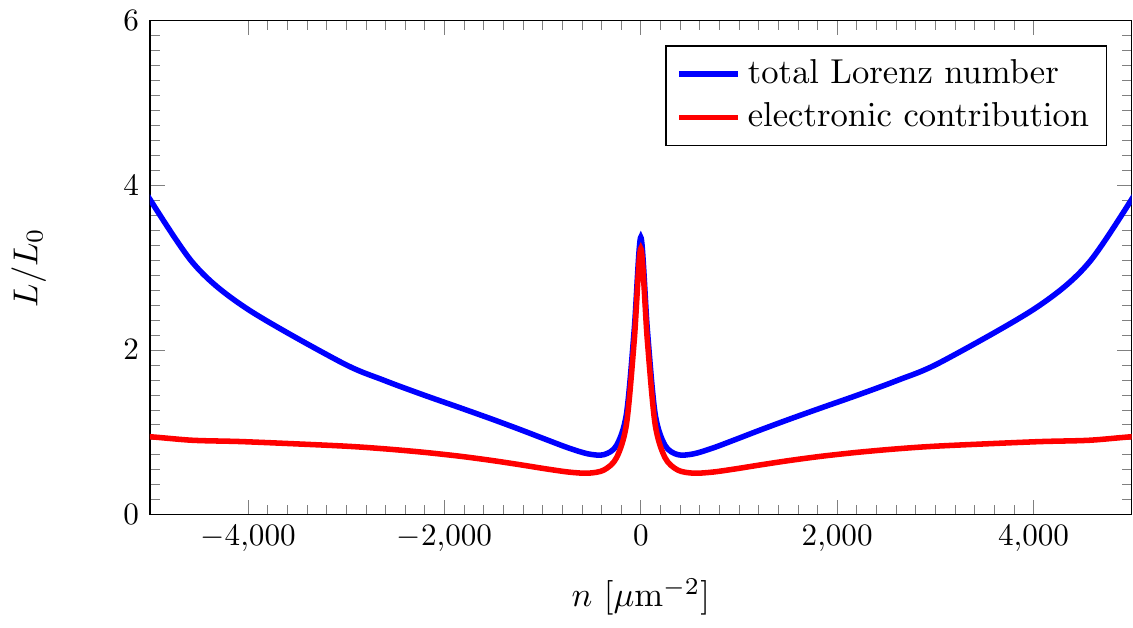}
\caption{Wiedemann-Franz ratio as a function of the charge carrier density. We observe two regions with enhancement: the well-known hydrodynamic regime in the vicinity of the Dirac point and at higher doping.}\label{fig:WF}
\end{figure}
\end{center}

\noindent{\it{Results:}} We have solved the coupled Boltzmann equations, Eq.~\eqref{eq:coupledboltz}, at and away from the Dirac point. This allows to access both the electrical and the heat conductivity, and consequently, the Wiedemann-Franz ratio. In the following we present two types of plots: (a) the conductivities as a function of the chemical potential (Fig.~\ref{fig:electricalconductivitymu}, Fig.~\ref{fig:heatconductivitymu},and Fig.~\ref{fig:WFmu}) and (b), more experimentally relevant, as a function of the electronic density (Fig.~\ref{fig:electricalconductivity}, Fig.~\ref{fig:heatconductivity},and Fig.~\ref{fig:WF}). In all plots, we have fixed the temperature to be $T=75$ K and the fine structure constant $\alpha$ to be $\alpha=0.36$. For disorder, we made the assumption that it is short-ranged and $4\pi \gamma^2=0.5$ (our main point here is not to connect to a specific experiment). In all plots, we plot the total conductivity including the plasmon contribution in blue and the electronic contribution only in red. Since the plasmons cannot make a direct contribution to the electrical conductivity, there is only a blue line in Fig.~\ref{fig:electricalconductivitymu} and Fig.~\ref{fig:electricalconductivity}. 
We observe that there is an enhancement of the thermal conductivity close to the Dirac point. This enhancement increases the expected violation of the Wiedemann-Franz law at the Dirac point. A bit more surprisingly, however, there is a sizeable and increasing enhancement of the thermal conductivity towards the Fermi liquid regime, {\it i.e.}, $\mu \gg T$.


It is tempting to attribute the growth of the plasmon contribution to its dispersion relation, Eq.~\eqref{eq:plasmondispersion}, which is $~\sqrt{\mu}$ for $\mu/T\gg 1$. However, this is not the sole reason for the increase: Phenomenologically, one expects the thermal conductivity to follow a Drude type expression $\kappa \propto \int d^2q \omega_p(\vec{q})\vec{v}^2_p(\vec{q}) \left( -\frac{\partial b}{\partial \omega_p(\vec{q})}\right) \tau_p(\vec{q},\mu)$, where $\tau_p(\vec{q},\mu)$ is a scattering time that comes from the solution of the Boltzmann equation. If we assume that $\tau_p(\vec{q},\mu)$ is constant as a function of $\vec{q}$ in the relevant momentum window, one ends up with $\kappa \propto \mu^0 \tau_p(0,\mu)$. Consequently, the effect appears to strongly depend on the scattering time, which is born out by an analysis of the scattering integral. 
To summarize, the main observation is the growing enhancement of the thermal conductivity due to plasmons in the region of $\mu/T >1$. It is important to note that the relaxation of the plasmons is due to the disorder in the fermionic sector. Momentum that is excited in the plasmon sector through the thermal gradient is transferred to the fermionic subsystem via inelastic scattering. There it is relaxed from the momentum conservation breaking disorder. 
It is worthwhile noting that the around $\mu/T \approx 2$ there is a suppression of the Lorenz ratio below 1. This seems to be a feature that is also encountered in experiments~\cite{Waissman}.

\noindent{\it{Conclusion and Outlook:}}  In this work we have analyzed the thermo-electric response of interacting two-dimensional Dirac systems at and away from the Dirac point. We have done so by deriving and solving coupled linearized Boltzmann equations for electrons, holes, and plasmons. At the Dirac point we find a moderate enhancement of the thermal conductivity due to plasmons, compared to the electronic contribution. However, away from the Dirac point, we find a strong enhancement of thermal transport due to plasmons. Compared to a conventional three dimensional metal, this is made possible by the undamped gapless nature of plasmons which have a square root dispersion, {\it i.e.}, $\omega \propto \sqrt{q}$. Consequently, this effect is special to two dimensions and is not expected to exist in three dimensions, in line with very early works~\cite{Pines&Bohm1952,Bohm&Pines1953}. The plasmon contribution to the heat conductivity and, connected to that, the violation of the Wiedemann-Franz law, increases as we tune into the Fermi liquid regime of a Dirac system. This suggests that the observed effect should also be observable in conventional degenerate two-dimensional metallic systems. While we do not expect a similar effect in three dimensional metals we also expect and enhancement close to the Dirac point of three dimensional Dirac-/Weyl-systems. 

The results presented here immediately provoke a series of questions: (i) How can the results be connected to the known results for the disordered two-dimensional Fermi liquid~\cite{CatelaniAleiner}? (ii) How can we model relaxation processes, such as disorder, for plasmons? (iii) Can this effect be observed in experiments? (iv) Can we find a unified hydrodynamic description in which all degrees of freedom enter on equal footing? The answers to some of the above questions are left for future study.


\noindent{\it{Acknowledgments:}} One of the authors (LF) acknowledges former collaborations and discussions with S. Sachdev, J. Schmalian, M. M\"uller, and Jonathan Lux as well as discussions with Andrew Lucas, Rembert Duine, Andrew Mitchell, Henk Stoof, and especially Jonah Waissman. KP thanks the Institute for the Promotion of Teaching Science and Technology (IPST) of Thailand for a Ph.D. fellowship.   This work is part of the D-ITP consortium, a program of the Netherlands Organisation for Scientific Research (NWO) that is funded by the Dutch Ministry of Education, Culture and Science
(OCW).



\pagebreak
\clearpage
\newpage
\widetext
\begin{center}
\textbf{\large Supplemental Material}
\end{center}
\setcounter{equation}{0}
\setcounter{figure}{0}
\setcounter{table}{0}
\setcounter{page}{1}
\makeatletter
\renewcommand{\theequation}{S\arabic{equation}}
\renewcommand{\thefigure}{S\arabic{figure}}
\renewcommand{\bibnumfmt}[1]{[S#1]}
\renewcommand{\citenumfont}[1]{S#1}

\section{Keldysh equations}

In the following we set up the Keldysh equations for describing transport phenomena in the coupled fermion-plasmon system. The procedure is akin a system of fermions coupled to phonons where drag effects have to be taken into account. 
Following standard procedure we parametrize the fermionic and bosonic Keldysh components as
\begin{eqnarray}
G^K&=& G^r \circ F-F \circ G^a \nonumber \\
D^K&=& D^r \circ B -B\circ D^a	
\end{eqnarray}
where $F$ and $B$ are hermitian matrices and $\circ$ denotes matrix multiplication in real space and time where $C=A\circ B$ corresponds to $C(\vec{x}_1,t_1,\vec{x}_2,t_2)=\int d\vec{x}' dt' A(\vec{x}_1,t_1,\vec{x}',t')B(\vec{x}',t',\vec{x}_2,t_2)$. Both $F$ and $B$ are, in thermal equilibrium, related to standard distribution function, where $F=1-2 f$ and $B=1+2 b$ with $f$ being the Fermi-Dirac distribution and $b$ the Bose-Einstein distribution. They obey the kinetic equation according to 
\begin{eqnarray}
\left[F,G_0^{-1}\right]^\circ_- &=& \Sigma^K- \left(\Sigma^r \circ F-F\circ \Sigma^a\right)	 \quad {\rm{and}}\nonumber \\  \left[B,D_0^{-1}\right]^\circ_- &=& e^2\left(\Pi^K- \left(\Pi^r \circ B-B\circ \Pi^a\right)\right)\;,
\end{eqnarray}
where the left-hand sides is a commutator involving the bare Green functions $G_0$ of the Dirac fermions and $D_0$ of the plasmons, whereas the right hand side is the so-called collision integral. We assume that $e^2\Pi$ is the self-energy of the bosons whereas $\Sigma$ is the self-energy of the fermions which we leave unspecified for the moment.
We are now going through a series of approximations which will eventually lead us to the Boltzmann equation. We start with a gradient expansion
\begin{eqnarray}\label{eq:kineticmoyal}
\left[F,G_0^{-1}\right]^\star_- &=& \left( \Sigma^K- \left(\Sigma^r \star F-F\star \Sigma^a\right)	\right)  \quad {\rm{and}}\nonumber \\   \left[B,D_0^{-1}\right]^\star_-  &=&  e^2 \left(\Pi^K- \left(\Pi^r \star B-B\star \Pi^a\right) \right)\;,
\end{eqnarray}
where we have introduced the Moyal product. It has to be interpreted in the following way: there are center of mass coordinates $X=(x_1+x_2)/2$ and $\tilde{T}=(t_1+t_2)/2$ as well as relative coordinates $x=x_1-x_2$ and $t=t_1-t_2$ (note that we introduce the notation ${\tilde{T}}$, here, to later distinguish it from the temperature $T$). Furthermore, we perform a Fourier transformation with respect to the relative coordinates, leading to $\vec{k}$ and $\omega$. The Moyal product then reads
\begin{eqnarray}
	C\left(X,{\tilde{T}},\vec{k},\omega \right)=A\left(X,{\tilde{T}},\vec{k},\omega \right) \star B\left(X,{\tilde{T}},\vec{k},\omega \right)
\end{eqnarray}
 with 
 \begin{eqnarray}
 \star = \exp \left[ \frac{i}{2} 	\left( \overleftarrow{\partial}_{\vec{X}}\overrightarrow{\partial}_{\vec{k}}-\overleftarrow{\partial}_{\tilde{T}} \overrightarrow{\partial}_\omega-\overleftarrow{\partial}_{\vec{k}}\overrightarrow{\partial}_{\vec{X}}  +\overleftarrow{\partial}_\omega^{\phantom{}} \overrightarrow{\partial}_{\tilde{T}} \right)\right]\;. \nonumber \\ 
 \end{eqnarray}
We perform a leading order expansion of both the left- and right-hand sides of Eq.~\eqref{eq:kineticmoyal}. 

\begin{eqnarray}
i\left( \partial_{\vec{X}}F \partial_{\vec{k}}\bar{G}^{-1}-\partial_{\vec{k}}F \partial_{\vec{X}}\bar{G}^{-1}-\partial_{\tilde{T}} F \partial_\omega \bar{G}^{-1} +\partial_\omega F \partial_{\tilde{T}} \bar{G}^{-1}\right)&=&  \Sigma^K- F\left(\Sigma^r - \Sigma^a\right) \nonumber \\ i\left(\partial_{\vec{X}}B \partial_{\vec{k}}\bar{D}^{-1}-\partial_{\vec{k}}B \partial_{\vec{X}}\bar{D}^{-1}-\partial_{\tilde{T}} B \partial_\omega \bar{D}^{-1} +\partial_\omega B \partial_{\tilde{T}} \bar{D}^{-1}\right)&=& e^2 \Pi^K- e^2 B\left(\Pi^r - \Pi^a\right)
\end{eqnarray}

where $\bar{G}^{-1}=G_0^{-1}-\Re{\Sigma^r}$ and $\bar{D}^{-1}=D_0^{-1}-e^2 \Re{\Pi^r}$. 

These two coupled equations constitute the basis of all further investigations. The next step towards the Boltzmann equation is to replace the function $F=1-2f$ and $B=1+2b$ with the respective distribution functions leading to

\begin{eqnarray}
i2\left(\partial_{\vec{k}}f \partial_{\vec{X}}G_0^{-1}-\partial_{\vec{X}}f \partial_{\vec{k}}G_0^{-1}+\partial_{\tilde{T}} f \partial_\omega G_0^{-1} -\partial_\omega f \partial_{\tilde{T}} G_0^{-1}\right) &=&  \Sigma^K- (1-2f)\left(\Sigma^r - \Sigma^a\right)\;, \nonumber \\ i2\left(\partial_{\vec{X}}b \partial_{\vec{k}}D_0^{-1}-\partial_{\vec{k}}b \partial_{\vec{X}}D_0^{-1}-\partial_{\tilde{T}} b \partial_\omega D_0^{-1} +\partial_\omega b \partial_{\tilde{T}} D_0^{-1}\right) &=& e^2 \Pi^K- e^2(1+2b)\left(\Pi^r - \Pi^a\right) \;.\nonumber \\	
\end{eqnarray}

In this paper we concentrate on heat and charge transport. For the left-hand sides of the kinetic equations this implies

\begin{eqnarray}
\partial_{\vec{k}}f \partial_{\vec{X}}G_0^{-1}-\partial_{\vec{X}}f \partial_{\vec{k}}G_0^{-1}+\partial_{\tilde{T}} f \partial_\omega G_0^{-1} -\partial_\omega f \partial_{\tilde{T}} G_0^{-1}&=& e \vec{E}\partial_{\vec{k}}f -v_F \vec{\sigma} \partial_{\vec{X}}f = e \vec{E}\partial_{\vec{k}}f -v_F \vec{\sigma}\cdot \partial_{\vec{X}}T \partial_T f \nonumber \\ \partial_{\vec{X}}b \partial_{\vec{k}}D_0^{-1}-\partial_{\vec{k}}b \partial_{\vec{X}}D_0^{-1}-\partial_{\tilde{T}} b \partial_\omega D_0^{-1} +\partial_\omega b \partial_{\tilde{T}} D_0^{-1}&=&2 \epsilon_0 \epsilon_r \frac{\vec{k}}{|\vec{k}|}\partial_{\vec{X}}b =2 \epsilon_0 \epsilon_r \frac{\vec{k}}{|\vec{k}|}\cdot \partial_{\vec{X}}T\partial_{T}b 
\end{eqnarray}

where $T$ is the temperature. The next step to convert this into a Boltzmann type equation is to perform the quasiparticle approximation. This is achieved by integrating over the spectral function. To that end we solve the Dyson equation to access the retarded Green functions $G^r$ and $D^r$. For $G^r$ suffices to state that the electrons of holes of graphene are well defined and we thus work with $G_0^r$ thereby disregarding corrections to infinitely long lived quasiparticles. For the plasmons, we use \begin{eqnarray}
	D^r(\vec{q},\omega)\approx \frac{1}{2 \epsilon_0 \epsilon_r}\frac{\omega^2}{q}	\frac{1}{(\omega+i0^+)^2-\omega_p(\vec{q})^2} 
\end{eqnarray}
with the plasmon dispersion  
\begin{eqnarray}
\omega_p(\vec{q})\approx  \sqrt{\alpha \frac{N}{2}k_B T v_F q \ln \left(2+2\cosh\left(\frac{\mu}{k_B T}\right)\right) }  \;,
\end{eqnarray}
as derived in the main text.
\section{Sources of relaxation}

The self-energy of the Dirac fermion consists of two parts: One due to interactions with the plasmons, another one due to scattering from impurities, to lowest order, is approximated as

\begin{eqnarray}
\Sigma^r(\omega,\vec{k})-\Sigma^a(\omega,\vec{k})&=& -2e^2\int \frac{d \nu}{2\pi} \int \frac{d^2q}{(2\pi)^2}\left[ {\rm{Im}} G^r(\omega+\nu,\vec{k}+\vec{q})\; D^K(-\nu,-\vec{q})  + G^K(\omega+\nu,\vec{k}+\vec{q})\;{\rm{Im}}D^r(-\nu,-\vec{q}) \right] \nonumber \\ &+&  \frac{\gamma_0^2}{2} \int \frac{d^2q}{(2\pi)^2} \hat{f}(-\vec{q})\left(G^r(\omega,\vec{k}+\vec{q})-G^a(\omega,\vec{k}+\vec{q})\right)\nonumber \\ \Sigma^K(\omega,\vec{k})&=& ie^2\int \frac{d \nu}{2\pi} \int \frac{d^2q}{(2\pi)^2}\left[ G^K(\omega+\nu,\vec{k}+\vec{q}) D^K(-\nu,-\vec{q})  -4 \;{\rm{Im}}G^r(\omega+\nu,\vec{k}+\vec{q})\;{\rm{Im}}D^r(-\nu,-\vec{q}) \right] \nonumber \\ &+& \frac{\gamma_0^2}{2}\int \frac{d^2q}{(2\pi)^2} \hat{f}(-\vec{q})G^K(\omega,\vec{k}+\vec{q})
\end{eqnarray}
where the first line in both cases accounts for scattering of plasmons whereas the second line accounts for disorder scattering. 

For the plasmons we have 

\begin{eqnarray}
\Pi^r(\omega,\vec{k})-\Pi^a(\omega,\vec{k})&=&N \int \frac{d\nu}{2\pi}\int \frac{d^2q}{(2\pi)^2}{\rm{tr}}\left( {\rm{Im}} G^r(\omega+\nu,\vec{k}+\vec{q})G^K(\nu,\vec{q})-G^K(\omega+\nu,\vec{k}+\vec{q}){\rm{Im}}G^r(\nu,\vec{q})\right)\nonumber \\ \Pi^K(\omega,\vec{k})&=&-\frac{i}{2} N \int \frac{d\nu}{2\pi}\int \frac{d^2q}{(2\pi)^2}{\rm{tr}}\left(G^K(\omega+\nu,\vec{k}+\vec{q})G^K(\nu,\vec{q})+4 \;{\rm{Im}}G^r(\omega+\nu,\vec{k}+\vec{q}){\rm{Im}}G^r(\nu,\vec{q})\right)
\end{eqnarray}

\section{Quasiparticle basis}

In order to arrive at the final Boltzmann equation we have to project the kinetic equations into the quasiparticle basis. This is straightforward for the plasmons, for the fermions we need a momentum-dependent rotation. To that end we consider the retarded part of the noninteracting fermionic Green function
\begin{eqnarray}
\left(G^r\right)^{-1}(\omega,\vec{k})	= \left(\omega+\mu \right)\mathds{1}-v_F k_x \sigma_x-v_F k_y \sigma_y
\end{eqnarray}
In order to project this onto the quasiparticle basis we need to diagonalize the Green function (or inverse Green function). The corresponding unitary transformation reads
\begin{eqnarray}
U^{-1}_{\vec{k}}&=&\frac{1}{\sqrt{2}k}\left(\begin{array}{cc} k_x-ik_y  & -k_x+ik_y \\ k & k \end{array} \right)	\nonumber \\ U^{\phantom{1}}_{\vec{k}}&=&\frac{1}{\sqrt{2}k}\left(\begin{array}{cc} k_x+ik_y  & k \\ -k_x-i k_y & k \end{array} \right)
\end{eqnarray}
with 
\begin{eqnarray}
\left(g^r\right)^{-1}(\omega,\vec{k})&=&U^{\phantom{1}}_{\vec{k}}\left(G^r\right)^{-1}(\omega,\vec{k})U^{-1}_{\vec{k}}\nonumber \\ &=&\left(\omega+\mu \right)\mathds{1}+v_F k \sigma_z \nonumber \\
\end{eqnarray}

\section{Coupled Boltzmann equations}

After having derived the Keldysh equations and specified the collision integral the last missing pieces towards the Boltzmann equation are a projection into the quasiparticle basis followed by an integration over the spectral functions. To that end we need the retarded part of the Dyson equation. For the plasmons, as discussed before, this reads
\begin{eqnarray}
(D^r)^{-1}(\omega,\vec{k})&=&(D_0^r)^{-1}(\omega,\vec{k})-e^2 \Pi^r (\omega,\vec{k}) \nonumber \\ &\approx & \frac{2 \epsilon_0 \epsilon_r k}{\omega^2}\left(\omega^2-\omega_p^2(\vec{k}) \right)	
\end{eqnarray}
whereas for the fermions we resort to the unperturbed propagator. We furthermore define the form factors
\begin{eqnarray}
M^{\lambda \lambda'}_{\vec{q},\vec{k}+\vec{q}}&=&\left( U^{\phantom{-1}}_{\vec{q}}U^{-1}_{\vec{k}+\vec{q}}\right)_{\lambda \lambda'}=\frac{1}{2}\left(1+\lambda \lambda' \frac{Q(K^\star+Q^\star)}{q|\vec{k}+\vec{q}|} \right)\nonumber \\ T^{\lambda \lambda'}_{\vec{q},\vec{k}+\vec{q}}&=&M^{\lambda \lambda'}_{\vec{q},\vec{k}+\vec{q}}M_{\vec{k}+\vec{q},\vec{q}}^{\lambda' \lambda}= \frac{1}{4}\left|\left(1+\lambda \lambda' \frac{Q(K^\star+Q^\star)}{q|\vec{k}+\vec{q}|} \right)\right|^2 \;.\nonumber \\
\end{eqnarray}
The poles of the Green function determine the dispersion $\epsilon_\lambda(\vec{k})=\lambda v_F k$. This allows to write the coupled Boltzmann equations as

\begin{eqnarray}
&&e\vec{E} \partial_{\vec{k}} f_{\lambda}(\vec{k})\delta_{\lambda \bar{\lambda}}-v_F \left(\hat{\vec{k}}\sigma_z-\hat{\vec{k}}\times  \hat{e}_z\sigma_y \right)_{\lambda \bar{\lambda}} \vec{\nabla} T \partial_T f_\lambda(\vec{k})\nonumber\\&&=4\pi v_F \alpha \int \frac{d^2q}{(2\pi)^2}\sum_{\lambda'=\pm} M^{\lambda \lambda'}_{\vec{k},\vec{k}+\vec{q}} M^{\lambda' \bar{\lambda}}_{\vec{k}+\vec{q},\vec{k}} \left( \delta \left(\epsilon_{\lambda'}(\vec{k}+\vec{q})-\epsilon_\lambda(\vec{k})+ \omega_p(\vec{q})\right)+\delta \left(\epsilon_{\lambda'}(\vec{k}+\vec{q})-\epsilon_\lambda(\vec{k})- \omega_p(\vec{q})\right)\right) \times \nonumber \\ &&\times \frac{\epsilon
_{\lambda'}(\vec{k}+\vec{q})-\epsilon_\lambda(\vec{k})}{q}\left[f_\lambda(\vec{k})\left(1-f_{\lambda'}(\vec{k}+\vec{q})\right)-b_{\epsilon_\lambda(\vec{k})-\epsilon
_{\lambda'}(\vec{k}+\vec{q})}(\vec{q})\left(f_{\lambda'}(\vec{k}+\vec{q})-f_\lambda(\vec{k})\right) \right] \nonumber \\ &&+ \frac{\gamma_0^2}{2\pi} \int \frac{d^2q}{(2\pi)^2} \hat{f}(-\vec{q})T_{\vec{k},\vec{k}+\vec{q}}^{\lambda \lambda}\delta_{\lambda \lambda'}\delta \left(\epsilon_\lambda (\vec{k})-\epsilon_{\lambda}(\vec{k}+\vec{q})\right)\left(f_\lambda(\vec{k})-f_{\lambda}(\vec{k}+\vec{q})\right) \nonumber \\ &&\frac{\vec{k}}{k^2}\vec{\nabla}T \omega_p(\vec{k})\left(\partial_T b_{\omega_p(\vec{k})}(\vec{k})-\partial_T b_{-\omega_p(\vec{k})}(\vec{k})\right)\nonumber \\ &&=4N\pi  v_F \alpha \sum_{\lambda,\lambda'=\pm}\int \frac{d^2q}{(2\pi)^2} M^{\lambda \lambda'}_{\vec{q},\vec{k}+\vec{q}} M_{\vec{k}+\vec{q},\vec{q}}^{\lambda' \lambda}\left( \delta \left(\epsilon_{\lambda'}(\vec{k}+\vec{q})-\epsilon_\lambda(\vec{q})+ \omega_p(\vec{k})\right)+\delta \left(\epsilon_{\lambda'}(\vec{k}+\vec{q})-\epsilon_\lambda(\vec{q})- \omega_p(\vec{k})\right)\right)\times \nonumber \\ &&\times \frac{\epsilon_{\lambda'}(\vec{k}+\vec{q})-\epsilon_\lambda(\vec{q})}{k}\left[ f_{\lambda'}(\vec{k}+\vec{q})\left(f_\lambda(\vec{q})-1\right)-b_{\epsilon_{\lambda'}(\vec{k}+\vec{q})-\epsilon_\lambda(\vec{q})}(\vec{k})\left(f_{\lambda'}(\vec{k}+\vec{q})-f_\lambda(\vec{q})\right) \right]\;.
\end{eqnarray}
In equilibrium we have
\begin{eqnarray}
f^0_\lambda(\vec{k})=\frac{1}{e^{(\epsilon_\lambda(\vec{k})-\mu)/T}+1} \quad {\rm{and}} \quad b^0_{\omega} (\vec{k})=\frac{1}{e^{\omega/T}-1}\;.
\end{eqnarray}

The second term in the l.h.s. of the first line corresponds to the Berry phase term which comes from the adiabatic projection into the quasiparticle basis. The fourth term is the thermal analogue of the Zitterbewegung. These terms makes no regular contribution in our calculation and are subsequently omitted.

\section{Linearized Boltzmann equation}
We then proceed to linearize the Boltzmann equations. To that end we introduce the parametrization
\begin{eqnarray}
f_\lambda(\vec{k})&=&f^0_\lambda(\vec{k})+\frac{1}{T^2}f^0_\lambda(\vec{k})\left(1-f^0_\lambda (\vec{k}) \right)\lambda v_F \hat{\vec{k}}	  \left(e\vec{E}  \chi^E_\lambda (k)+ \vec{\nabla}T   \chi^T_\lambda (k) \right) \nonumber \\ b_{\omega_p(\vec{k})} (\vec{k})&=& b^0_{\omega_p(\vec{k})}(\vec{k})+ \frac{1}{T^2} b^0_{\omega_p(\vec{k})}(\vec{k})\left( 1+ b^0_{\omega_p(\vec{k})}(\vec{k})\right) v_F \hat{\vec{k}}  \left(e\vec{E}  \phi^E (k)+ \vec{\nabla}T   \phi^T (k) \right)
\end{eqnarray}

Using the linearization and neglecting the Berry phase as well as the off-diagonal contribution we obtain

\begin{eqnarray}
 	&& -\lambda e \frac{v_F}{T} \hat{\vec{k}}\cdot\vec{E}  f^0_{\lambda}(\vec{k})\left(1-f^0_{\lambda}(\vec{k}) \right)-\lambda \frac{v_F}{T} \hat{\vec{k}} \vec{\nabla} T \frac{\epsilon_\lambda(\vec{k})-\mu}{T} f^0_{\lambda}(\vec{k})\left(1-f^0_{\lambda}(\vec{k}) \right)  \nonumber\\&&=\alpha\frac{4\pi v_F^2}{T^2}  \int \frac{d^2q}{(2\pi)^2}\sum_{\lambda'=\pm} M^{\lambda \lambda'}_{\vec{k},\vec{k}+\vec{q}} M^{\lambda' \lambda}_{\vec{k}+\vec{q},\vec{k}} \left( \delta \left(\epsilon_{\lambda'}(\vec{k}+\vec{q})-\epsilon_\lambda(\vec{k})+ \omega_p(\vec{q})\right)+\delta \left(\epsilon_{\lambda'}(\vec{k}+\vec{q})-\epsilon_\lambda(\vec{k})- \omega_p(\vec{q})\right)\right) \times \nonumber \\ &&\times \frac{\epsilon_{\lambda'}(\vec{k}+\vec{q})-\epsilon_\lambda(\vec{k})}{q}f^0_\lambda(\vec{k})\left(1-f^0_{\lambda}(\vec{k})\right)\left(1-f^0_{\lambda'}(\vec{k}+\vec{q})+ b^0_{\epsilon_\lambda(\vec{k})-\epsilon_{\lambda'}(\vec{k}+\vec{q})}(\vec{q})\right)\times \nonumber \\ &&\times \left[e\vec{E} \left(\lambda \frac{\vec{k}}{k} \chi_\lambda^E-\lambda' \frac{\vec{k}+\vec{q}}{|\vec{k}+\vec{q}|}\chi^E_{\lambda'}(|\vec{k}+\vec{q}|)+\frac{\vec{q}}{q} \phi^E(q) \right)+\vec{\nabla} T \left(\lambda \frac{\vec{k}}{k} \chi_\lambda^T-\lambda' \frac{\vec{k}+\vec{q}}{|\vec{k}+\vec{q}|}\chi^T_{\lambda'}(|\vec{k}+\vec{q}|)+\frac{\vec{q}}{q} \phi^T(q) \right) \right] \nonumber \\ &&+ \frac{\gamma_0^2 v_F}{2\pi T^2} \int \frac{d^2q}{(2\pi)^2} \hat{f}(-\vec{q})T_{\vec{k},\vec{k}+\vec{q}}^{\lambda \lambda}\delta_{\lambda \lambda'}\delta \left(\epsilon_\lambda (\vec{k})-\epsilon_{\lambda}(\vec{k}+\vec{q})\right)f^0_\lambda(\vec{k})\left( 1-f^0_{\lambda}(\vec{k})\right) \times \nonumber \\ && \times \left[e\vec{E} \left(\lambda \frac{\vec{k}}{k} \chi_\lambda^E-\lambda' \frac{\vec{k}+\vec{q}}{|\vec{k}+\vec{q}|}\chi^E_{\lambda'}(|\vec{k}+\vec{q}|) \right)+\vec{\nabla} T \left(\lambda \frac{\vec{k}}{k} \chi_\lambda^T-\lambda' \frac{\vec{k}+\vec{q}}{|\vec{k}+\vec{q}|}\chi^T_{\lambda'}(|\vec{k}+\vec{q}|) \right) \right]  \nonumber \\ &&\frac{\vec{k}}{k^2}\vec{\nabla}T \frac{\omega_p^2(\vec{k})}{T^2}b^0_{\omega_p(\vec{k})}(\vec{k})\left( 1+ b^0_{\omega_p(\vec{k})}(\vec{k})\right)\nonumber \\ &&=\alpha 4N\pi  \frac{v_F^2}{T^2}  \sum_{\lambda,\lambda'=\pm}\int \frac{d^2q}{(2\pi)^2} M^{\lambda \lambda'}_{\vec{q},\vec{k}+\vec{q}} M_{\vec{k}+\vec{q},\vec{q}}^{\lambda' \lambda}\left( \delta \left(\epsilon_{\lambda'}(\vec{k}+\vec{q})-\epsilon_\lambda(\vec{q})+ \omega_p(\vec{k})\right)+\delta \left(\epsilon_{\lambda'}(\vec{k}+\vec{q})-\epsilon_\lambda(\vec{q})- \omega_p(\vec{k})\right)\right)\times \nonumber \\ &&\times \frac{\epsilon_{\lambda'}(\vec{k}+\vec{q})-\epsilon_\lambda(\vec{q})}{k}f^0_\lambda(\vec{q})\left( 1-f^0_\lambda(\vec{q})\right) \left(f^0_{\lambda'}(\vec{k}+\vec{q})+b^0_{\epsilon_{\lambda'}(\vec{k}+\vec{q})-\epsilon_\lambda(\vec{q})}(\vec{k})\right) \times \nonumber \\ && \times \left[e\vec{E} \left(\lambda \frac{\vec{q}}{q} \chi_\lambda^E-\lambda' \frac{\vec{k}+\vec{q}}{|\vec{k}+\vec{q}|}\chi^E_{\lambda'}(|\vec{k}+\vec{q}|)+\frac{\vec{k}}{k} \phi^E(k) \right)+\vec{\nabla} T \left(\lambda \frac{\vec{q}}{q} \chi_\lambda^T-\lambda' \frac{\vec{k}+\vec{q}}{|\vec{k}+\vec{q}|}\chi^T_{\lambda'}(|\vec{k}+\vec{q}|)+\frac{\vec{k}}{k} \phi^T(k) \right) \right]
\end{eqnarray}

Our ansatz for the deviation from equilibrium reads
\begin{eqnarray}
\chi^E_\lambda (k)&=&a^E_{0,\lambda}+a^E_{1,\lambda} k=a^E_{0,\lambda}|E,0,\lambda\rangle +a^E_{1,\lambda} |E,1,\lambda \rangle \nonumber \\ 	\chi^T_\lambda (k)&=&a^T_{0,\lambda}+a^T_{1,\lambda} k = a^T_{0,\lambda}|T,0,\lambda\rangle +a^T_{1,\lambda} |T,1,\lambda \rangle\nonumber \\ \phi^E(k)&=&b_0^E+b^E_1 k =b_0^E |E,0\rangle + b^E_1 |E,1\rangle \nonumber \\ \phi^T(k)&=&b^T_0+b^T_1 k=b^T_0|T,0\rangle+b^T_1|T,1\rangle \;. 
\end{eqnarray}
One can rewrite the Boltzmann equations in a more compact form as
\begin{eqnarray}\label{eq:ansatz}
|Df,E,\lambda \rangle+|Df,T,\lambda \rangle &=&|I_{\rm{coll}},E\rangle+|I_{\rm{coll}},T\rangle	\nonumber \\ |Db,T\rangle &=& |\tilde{I}_{\rm{coll}},E\rangle+|\tilde{I}_{\rm{coll}},T\rangle \;.
\end{eqnarray}
To determine the expansion coefficients in Eq.~\eqref{eq:ansatz} we define a scalar product according to
\begin{eqnarray}
\langle f| g\rangle = \int \frac{d^2k}{(2\pi)^2} f(k)\frac{\vec{k}}{k} g(\vec{k})	\;.
\end{eqnarray}
This allows to convert the linearized Boltzmann equations into a linear algebra problem which we can solve for $a_{0/1,\lambda}^{E/T}$ and $b_{0/1}^{E/T}$.

\end{document}